\documentclass[aps,twocolumn,showpacs,amsmath]{revtex4-1}
\usepackage{graphicx}
\usepackage{amsmath}

\begin{document}

\title{The Schr\"odinger equation for general non-hermitian quantum system}

\author{Ye Xiong}
\email{xiongye@njnu.edu.cn}
\affiliation{Department of Physics and Institute of Theoretical Physics
  , Nanjing Normal University, Nanjing 210023, P. R. China \\
   National Laboratory of Solid State Microstructures, Nanjing
  University, Nanjing 210093, P. R. China}

\author{Peiqing Tong}
\email{pqtong@njnu.edu.cn}
\affiliation{Department of Physics and Institute of Theoretical Physics
  , Nanjing Normal University, Nanjing 210023,
P. R. China \\
Jiangsu Key Laboratory for Numerical Simulation of Large
  Scale Complex Systems, Nanjing Normal University, Nanjing 210023,
P. R. China}

\begin{abstract}
  We derive a new time-dependent Schr\"odinger equation(TDSE) for
  quantum models with non-hermitian Hamiltonian. Within our theory, the TDSE is
  symmetric in the two Hilbert spaces spanned by the left and the right
  eigenstates, respectively. The physical quantities are also
  identical in these two spaces. Based on this TDSE, we show
  that exchanging two quasi-particles in a non-hermitian model can generate
  arbitrary geometric phase. The system can also violate
  the Lieb-Robinson bound in non-relativistic quantum mechanics so
  that an action in one place will immediately cause a change in the
  distance. We show that the above two surprising behaviors can
  also appear in anyonic model, which makes us propose that the 
  non-hermitian single particle model may possess
  many common features with anyonic model. 
\end{abstract}

\maketitle

In quantum mechanics, the time-dependent Schr\"odinger equation(TDSE):
\begin{equation}
i \hbar \partial_t |\Psi\rangle =\hat H
|\Psi \rangle,
\label{eq1}
\end{equation}
is crucially important in determining the evolution of the
wave-function $|\Psi\rangle$. But how will this equation be changed
when the Hamiltonian $\hat H$ becomes non-hermitian(NH) and explicitly
time-dependent, is still under debates\cite{0305-4470-39-29-018, Faria2007, Mostafazadeh2007208,
Znojil1, Mostafazadeh1, Znojil2,Mostafazadeh2,PhysRevD.78.085003,
Gong2013, PhysRevA.92.032106, PhysRevA.93.042114, PhysRevA.94.042128}. 

Although the discussions on the possibility of NH Hamiltonian 
started more than half a century
ago\cite{RevModPhys.29.269}, this vital gap has not been filled. The
situation does not get better even when a specific kind of NH systems,
$\mathcal{PT}$-symmetric systems, has been systematically studied both
theoretically\cite{Bender1998, Bender2002, Mostafazadeh2002, Cui2012,
Lee2014, Malzard2015, Medvedyeva2016, Amir2015, Dattoli1990,
Mehri-Dehnavi2008, Muller2009a, Gania2010, Liang2013, Cui2014a, Lee2014b,
Bender2014a, Shah2015, Nixon2015, Xiong1}
and experimentally\cite{0305-4470-38-9-L03, PhysRevLett.100.103904,
  PhysRevLett.103.093902, PhysRevA.82.031801, PhysRevLett.103.123601,
  Ruter2010, PhysRevA.82.043803, PhysRevLett.106.213901,
  PhysRevA.85.050101, PhysRevLett.110.234101,PhysRevLett.113.023903,
  Regensburger2013, Zhang2016g, Hodaei2014, PhysRevX.4.031042,
  Fleury2015,Chen2016}. The obstacle lies in the fact that NH
Hamiltonian $\hat H$ has the right eigenstates $|n \rangle$ and the left
eigenstates $\langle \langle n|$: 
\begin{eqnarray}
  \hat H |n \rangle = E_n |n \rangle, \\
  \langle \langle n| \hat H = E_n \langle \langle n|, 
  \label{eq2}
\end{eqnarray}
which are not conjugate to each other: $\langle \langle n| \ne (|n
\rangle)^\dagger$. So a NH Hamiltonian induces multiple mappings into
two Hilbert spaces spanned by $|n \rangle$ and $|n \rangle \rangle$,
respectively\cite{Scholtz1992,Znojil1}. Here $\langle n |$ and $|n \rangle \rangle$ are the
hermitian conjugate of $|n \rangle$ and $\langle \langle n |$. In
either Hilbert space, the inner product of wave functions,
$\langle \Psi | W | \Psi
\rangle$ and $\langle \langle \Psi | W' | \Psi \rangle \rangle$, where
$W$ and $W'$ are called the metric operators in the Hilbert spaces\cite{Scholtz1992,
Mostafazadeh2002, Bender2002, Mocal3, Czech1,Mocal4,Mocal5}, is also
distinct from $\langle \Psi | \Psi \rangle$ in the hermitian case. As the
constraints on the metric are only positive-definite and hermitian, 
different TDSEs are
formulated by choosing different metrics. For instance, Faria chose a
time-independent metric\cite{0305-4470-39-29-018}, Fring used the metric
obeying his equation Eq. (2.6) in Ref. \onlinecite{PhysRevA.93.042114} while
Gong chose a metric defined by the eigenstates\cite{Gong2013}. Besides
that, physicists lack a suitable system to examine these equations.
Although the $\mathcal{PT}$ models have been realized in experiments, they
cannot be used to inspire the problem for the two reasons. One is that the
Hamiltonians in these systems are time-independent. The other is
more essential: the $\mathcal{PT}$ systems are mostly governed by the predefined
fundamental rules instead of the NH Hamiltonian, e.g., Maxwell equations
in optics \cite{PhysRevLett.100.103904} or Newton's law in classical
acoustic systems \cite{PhysRevX.4.031042}. So the requirements on the
quantum NH TDSE, such as the evolution should be unitary, may not subject to
the experimental systems at all.

In this article, by assuming that the damping rates, which are associated with the
imaginary parts of the eigenvalues of the Hamiltonian, are physically
meaningful, we propose that the TDSE for a general NH system should read
as
\begin{equation}
  i\hbar \partial_t | \Psi \rangle = [\frac{ {\tilde W}^{-1} {\hat
  H}^\dagger \tilde W
+\hat H }{2}-\frac{i\hbar}{2}{\tilde W}^{-1} \dot{\tilde W}] |\Psi \rangle, 
  \label{eq3}
\end{equation}
where the instantaneous metric $\tilde W$ is defined by the eigenstates of the
Hamiltonian and $\dot{\tilde W}$ refers to its time derivative.
The Born's representation on the wave-function, the probability
of finding the quasi-particle at the state $i$, is redefined by a metric
connection
$W$ that connecting the instantaneous metrics $\tilde W$ from
initial time $t=0$. We will show that, due to the extra $\frac{i\hbar}{2}{\tilde W}^{-1}
\dot{\tilde W}$ term, a quench in NH quantum
system is significantly different from that in the hermitian case.
By solving the differential equation,
the wave-function infinitely after the quench is distinct from the one
infinitely before the quench. This effect is
``an action in distance'' that violates the Lieb-Robinson bound(LRB).
Although a similar result has been given in Ref. \onlinecite{Lee2014}, we
want to emphasize that the starting point of the discussions are totally
different: one is from the traditional TDSE in Eq. \ref{eq1} and the
other is from the new TDSE in Eq. \ref{eq3}. We also find that, by
exchanging two quasi-particles adiabatically in such a
NH system, the accumulated geometric phase 
can be different from $\pi$, a phenomenon similar to exchanging
two anyons\cite{Laughlin1983, Arovas1984, Graner1992}. As anyonic
system may also possess a similar ``an action in distance''
during quench, we reveal that a NH quantum system {\it without} many-body
interaction can share many features with anyonic system.

{\it The time-dependent Schr\"odinger equation for nonhermitian
system.---} In hermitian quantum system, the Hamiltonian $\hat H$ plays
dual roles. On the one side, it determines the evolution of wave function
through the TDSE in Eq. \ref{eq1}. On the other side, the eigenvalues of the Hamiltonian
are physical meaningful: they are the energy levels of the
system. But in NH quantum system, it is commonly accepted that such
duality is destroyed \cite{0305-4470-39-29-018, Gong2013,
PhysRevA.93.042114}. So there are two options: one is by retaining the
TDSE in Eq. \ref{eq1} while abandoning the energy representation of the
Hamiltonian and the other is vice versa.  

Several authors have adopted the latter option. For
instance, by normalizing the total probability to unity, 
Wieser wrote down a new TDSE\cite{Wieser2013},
\begin{equation}
  i\hbar \partial_t |\Psi(t)\rangle = [\hat H + \langle \Psi(t) |
  \frac{ ({\hat H}^\dagger-{\hat H} )}{2} |\Psi(t) \rangle ] |\Psi
  \rangle.
  \label{eq4}
\end{equation}
He had successfully derived the semi-classical Landau-Lifshitz equation from
this fully quantum equation. Although we does not agree with him on this
equation in general because the metric has not been considered entirely, 
Wieser's work implies that the imaginary parts of the eigenvalues of the
Hamiltonian,
$\Im(E_n)$, are characterizing the damping rates of the corresponding
eigenstates. This makes us adopt the same option by keeping the eigenvalues of
the Hamiltonian as physically meaningful and modifying TDSE. Besides
that, his work also indicates that the overall evolution is nonlinear in
the presence of damping. This makes us think how to unify the TDSE with 
the superposition principle in the NH quantum mechanics.

Gong, in Ref. \onlinecite{Gong2013}, proposed another TDSE reading as 
\begin{equation}
  i \hbar \partial_t |\Psi \rangle = (\hat H - \frac{i\hbar}{2}W^{-1}
  \dot{W}) |\Psi \rangle,
  \label{eq5}
\end{equation}
The effect of the metric has been considered, but the equation can only
subject to a special kind of NH models with
conserved PT symmetry, $\Im(E_n)=0$. Inspired by this work, we
build up the TDSE in Eq. \ref{eq3} that can be applied to any NH system.
And our equation can be retrieved to Eq. \ref{eq5} in $\mathcal{PT}$-symmetric
cases and to Eq. \ref{eq4} when the metric is a unit matrix.

As there is biorthonormal relation for the left eigenstates $\langle
\langle n|$ and the right eigenstates $|n \rangle$ \cite{Mocal6,
Mostafazadeh2002}, $\langle \langle n |
m \rangle = \delta_{nm}$ after proper normalization, we can define the instantaneous metric as 
\begin{equation}
\tilde W =\sum_n \tilde W_n , \; \tilde W_n=
|n \rangle \rangle  \langle \langle n |.
\label{Weq}
\end{equation}
Here we have omitted the time-dependent notation for brevity.  
Each individual $\tilde W_n$ can help to determine the components of the wave
function $|\Psi \rangle$ in the Hilbert space spanned by $|n\rangle$,
$|c_n|^2 \propto \langle \Psi | \tilde W_n | \Psi \rangle$. The damping
(with $\Im(E_n)<0$) and inflation (with $\Im(E_n)>0$) of the basis
in this space can be absorbed by the definition of a metric connection $W$. This
is our first ansatz: if the wave-function $|\Psi \rangle$ is
found, we can determine its components in each eigenstate 
by $|c_n(t)|^2 = \frac{1}{A(t)} \langle \Psi | W_n | \Psi \rangle$,
where $W_n(t) = \tilde W_n(t) e^{\int_{t_0}^{t} 2\Im(E_n(t'))dt'}$, $t_0$
is the initial time and the normalizing factor $A(t)$ is also a function
of $|\Psi\rangle$, $A(t)= \sum_n
\langle \Psi | W_n | \Psi \rangle$. The observation of an observable
operator $\hat O$ is $o= \frac{1}{A} \langle \Psi | \eta^\dagger \hat O
\eta | \Psi \rangle$\cite{Gong2013,PhysRevA.93.042114} with the metric connection explicitly written as
\begin{equation}
  W = \eta^\dagger \eta =\sum_n |n(t) \rangle \rangle e^{\int_{t_0}^{t}
2\Im(E_n(t'))dt'} \langle \langle n(t) |.
  \label{eq6}
\end{equation}

As the damping has been absorbed in the metric connection, we can 
impose the unitary condition on the wave function when acting with
instantaneous metric $\tilde W$, 
\begin{equation}
  \frac{d}{dt} (\langle \Psi | \tilde W |\Psi \rangle) =0.
  \label{eq7}
\end{equation}
After that, we employ the second ansatz: the TDSE should reads as
$i\hbar \partial_t
|\Psi \rangle = (\hat H + \Lambda) |\Psi \rangle$, where $\hat H$ is the
Hamiltonian and $\Lambda$ can exchange with $\tilde W$ like $\tilde W
\Lambda = \alpha \Lambda^\dagger \tilde W$. Here $\alpha$ is a number to
be determined self-consistently. After substituting the above ansatz to
the unitary condition in Eq. \ref{eq7}, one can find $\alpha=-1$
and the TDSE in Eq. \ref{eq3}.
   
We have a few remarks on the equations. Firstly, they implies that the
equations of motion for the wave functions $|\Psi \rangle$ and $|\Psi
\rangle\rangle$, that lie in the Hilbert spaces spanned by $|n\rangle$
and $|n\rangle\rangle$ respectively, are symmetric. From $|\Psi
\rangle\rangle= \tilde W |\Psi\rangle$, we can find
\begin{equation}
i\hbar \partial_t | \Psi \rangle\rangle = [\frac{ {\tilde W} {\hat
H} {\tilde W}^{-1}
+\hat H^\dagger }{2}-\frac{i\hbar}{2}{\tilde W} \frac{d{\tilde
W}^{-1}}{dt}]
|\Psi \rangle\rangle.
\label{eq8}
\end{equation}
This equation is symmetric to Eq. \ref{eq3} by exchanging $\hat H$ with $\hat
H^\dagger$ and $\tilde W$ with ${\tilde W}^{-1}$ (here the metric in the
latter Hilbert space has changed to ${\tilde W}^{-1}$). This is sound
because, in principle, the two Hilbert spaces spanned 
by the right eigenstates and the left eigenstates of 
a matrix, are equally weighted. If one insists on the traditional TDSE
like Eq. \ref{eq1}, such symmetry 
will be destroyed. Secondly, although Eq. \ref{eq3} seems linear, the overall
system is nonlinear because the calculations of physical quantities are depending
on the metric connection $W$. One can catch this easily in such a
demo model. Let $\tilde W(t) =1$ at any time and $\hat H = |1\rangle E_1
\langle 1|+|2\rangle E_2 \langle 2|$. Here the left eigenstate
$\langle \langle n|$ retrieves to the hermitian conjugate of the right
eigenstate $|n\rangle$ because $\tilde W=1$. For an initial state $|\Psi
\rangle = c_1 |1 \rangle + c_2 |2\rangle$, $|c_i(t)|^2$ will change with
time as $\frac{1}{A(t)} |c_i|^2 \exp{\int_0^t dt' 2\Im(E_i(t'))}$ where
the normalized factor $A(t)= |c_1(t)|^2+ |c_2(t)|^2$ is dependent
on $|c_i(t)|^2$ nonlinearly when $\Im(E_i(t'))$ is not zero. Such kind
of equation is equivalent to Eq. (4) in Wieser's letter\cite{Wieser2013}. 
So we can claim that our equation can be retrieved to Wieser's one when the instantaneous
metric $\tilde W(t)$ is a unit matrix all the time. Thirdly, to
calculate the observation of an operator $\hat O$, our equation relies
on $\eta$, which is related with the metric connection $W$ by
$W=\eta^\dagger \eta$. But a given $W$
can map to infinite $\eta$s because by unitary transforming  
$\eta \to U\eta$, $\eta^\dagger\eta$ is still $W$. We would like to emphasize
that such a transformation is associated with an unitary translation in the
Hilbert spaces so that the operator $\hat O$ should also be changed as $
\hat O \to U \hat O U^\dagger$. So the observation values are not modified by such
transformation. Fourthly, when $\mathcal{PT}$-symmetry is reserved,
${\tilde W}^{-1} \hat H^\dagger \tilde W =\hat H$ and $\Im(E_i)=0$.
Our equation will retrieve to Gong's equation in Eq. \ref{eq5}. Fifthly,
the extra ${\tilde W}^{-1} \dot{\tilde W}$ term can trigger many interesting
phenomena. We will only cover two of them in this article. One is that
$\tilde W$ can be discontinuous when quenching a Hamiltonian.
This will introduce a $\delta$-like function on the right-hand side of Eq.
\ref{eq3} because of the presence of $\dot{\tilde W}$. By solving this
first-order differential equation, one will find that the wave function is
also discontinuous. The other interesting phenomenon is
that the expression of geometric phase is also different from that in the
traditional quantum mechanics.

{\it An action in distance.}--- 
In hermitian quantum system, a quench is usually
employed by suddenly changing the Hamiltonian from $\hat H_0$ to $\hat
H_1$ at time $t=0$. As the time interval of the quench is absolutely
zero, the wave functions before and after the quench,
$|\Psi(t=0^-)\rangle$ and $|\Psi(t=0^+)\rangle$, are identical. But 
this is not the case for NH system.

As the instantaneous metrics, $\tilde W_- = \tilde W(t=0^-)$ and $\tilde
W_+=\tilde W(t=0^+)$
are different, one cannot solve the TDSE in Eq. \ref{eq3} at $t=0$
without properly regulating the differential equation. We can
actually solve this by employing the unitary condition in Eq. \ref{eq7}
directly, which gives
\begin{equation}
\langle \Psi(0^-) |\tilde W_-| \Psi(0^-) \rangle = 
\langle \Psi(0^+) |\tilde W_+| \Psi(0^+) \rangle.
\end{equation}
By denoting the evolution as $|\Psi(0^+) \rangle =\hat L | \Psi(0^-)
\rangle$, we find that the solution to the above equation is $\hat L = 
(\sqrt{\tilde W_+})^{-1}U \sqrt{\tilde W_-}$,
where $U$ is a unitary matrix. Due to the anti-symmetric relation
$\tilde W \Lambda = - \Lambda^\dagger \tilde W$, we also have $\tilde
W_+ \hat L = \hat L^\dagger \tilde W_+$. After some algebra, we find $U$
can be determined by
\begin{eqnarray}
  U (\sqrt{\tilde W_-} {\tilde W_+}^{-1} \sqrt{\tilde W_-}) U^\dagger =  
  \\ \nonumber
  (\sqrt{\tilde W_+})^{-1}{\tilde W_-} (\sqrt{\tilde
  W_+})^{-1}.
  \label{eqU}
\end{eqnarray}
Here $\sqrt{\tilde W_{\pm}}$ is still hermitian by making the square
root of the eigenvalues of $\tilde W_{\pm}$.

When representing such equation in real space, we realize that LRB
should be violated. LRB states that one can not measure the signal of
change outside a cone around the source of change (at where the quench
takes place). The slope of the cone is referring to the maximal speed of the
signal in the system. But as $\hat L$ is
off-diagonal in general, it is implied that a
quench at a source can immediately affect the wave function far away
from it. This is actually an effect of ``an action in distance''. One should note that
LRB in non-relativistic quantum system is not protected by any basic
principle so that its violation does not implies the fault of the relativistic
principle.

{\it The geometric phase.}---
As the wave function is evolving according to the new TDSE, the form of the
geometric phase is also changed. We suppose that the evolution is so
slow that the jumps between the eigenstates are ignored. So the evolution of
eigenstates can be written as $|\Psi_n(t) \rangle = c_n(t) | n(t)
\rangle$, where $|n(t)\rangle$ is the instantaneous eigenstate. Beside
the dynamical phase, $c_n(t)$ also possesses a geometric factor like
$\exp(i\gamma_n)$, where $\gamma_n$ is the geometric phase. After
substituting the above wave function to the TDSE, we get the 
geometric phase as
\begin{equation}
  \dot \gamma_n = i [\langle n | \tilde W | \dot n \rangle +
  \frac{1}{2} \langle n |\dot{\tilde W} | n \rangle]. 
  \label{eq10}
\end{equation}
This form of geometric phase is the same as that in Ref. 
\onlinecite{Gong2013}, because the two TDSEs are only distinguished in
the dynamical part. One can also express the geometric phase by the left and the
right eigenstates like $\dot \gamma_n = \frac{i}{2}[\langle
\langle n | \dot n \rangle + \langle n | \dot n \rangle \rangle]$.
Similarly, when expanding the wave function in the Hilbert space spanned
by the left eigenstates $|n \rangle \rangle$, we can find that the
geometric phase in this space is the same. This makes us raise the
third antatz which is used to check the correction of equations: In NH
system, physical quantities should be identical and the formulas should
be symmetric in
either Hilbert space spanned by the left or by the right eigenstates.

Now we discuss the geometric phase by exchanging two quasi-particles in
a NH system. We start with a $2\times 2$ Hamiltonian so that there are
only two eigenstates denoted by $1$ and $2$ respectively. The
eigenstates are represented by the points in the Bloch sphere, $|1
\rangle = \frac{1}{A}\begin{pmatrix} \cos(\theta/2) \\ \sin(\theta/2) e^{i\phi}
\end{pmatrix}$ and $|1
\rangle \rangle= \begin{pmatrix} \cos(\theta'/2) \\ \sin(\theta'/2)
  e^{i\phi'}
\end{pmatrix}$, where
$A=\cos(\theta/2)\cos(\theta'/2)+\sin(\theta/2)\sin(\theta'/2)
e^{i(\phi-\phi')}$ is the factor to normalize $\langle \langle 1 | 1
\rangle$ and $(\theta,\phi)$ are the polar angles. One should note
that $|1\rangle$ and $|1 \rangle \rangle$ can refer to different
points in the Bloch sphere in NH case. The eigenstate $|2\rangle$, which should be
orthogonal to $|1 \rangle \rangle$, is get by replacing $\theta'
\to \pi - \theta'$ and $\phi' \to \phi'+\pi$. Similarly $|2 \rangle
\rangle$ can also be found from a similar replacement on $\theta$ and $\phi$.
Here the normalization factor $A$ for $|2\rangle$ is different from that for $|1\rangle$. 
 
\begin{figure}[htp]
  \centering
  \includegraphics[width=0.5\textwidth]{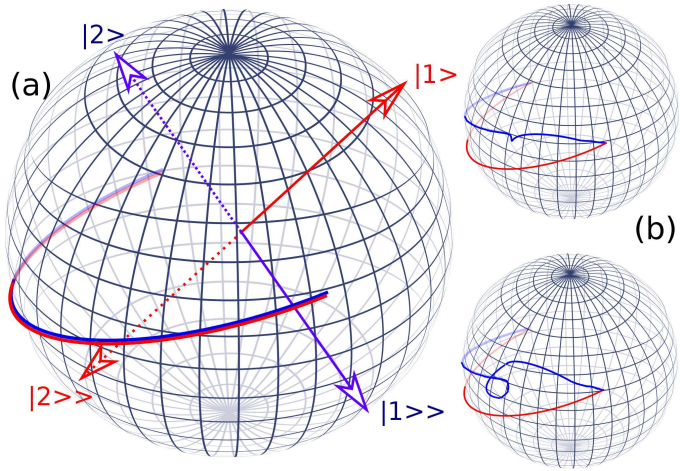}
  \caption{(a) The eigenstates $|1\rangle$(solid red),
  $|2\rangle$(dashed blue), $|1\rangle \rangle$(solid blue)
  and $|2\rangle \rangle$(dashed red) are presented in the Bloch sphere. 
  The orthogonal restrictions $\langle 1| 2\rangle \rangle=0$ and $\langle
  2| 1\rangle \rangle=0$ are satisfied. For
the hermitian Hamiltonian, the traces of exchanging the states $1$ and $2$
are coincided for the left and the right states. Here we will only show
the traces for $|1\rangle$ and $|1\rangle\rangle$ for the sake of
clarify. (b) In NH model, the traces for $|1\rangle$ and $|1\rangle
\rangle$ can be different. In the upper panel, the traces do not
have loop and the geometric phase is
still $\pi$. But in the lower panel, when the traces have loop, the
geometric phase can be any values. For the traces shown in
the panel, the geometric phase is $0.92\pi$.}
  \label{fig1}
\end{figure}

As schematically showed in Fig. \ref{fig1} (a), when the traces for
$|1\rangle$ and $|1 \rangle\rangle$ coincide, the model will retrieve to
the hermitian case. In that case, when two fermions at the states $|1
\rangle = \frac{\sqrt{2}}{2} \begin{pmatrix} 1 \\ 1 \end{pmatrix}$ and
$|2 \rangle =  \frac{\sqrt{2}}{2} \begin{pmatrix} 1 \\ -1 \end{pmatrix}$
are exchanged adiabatically along the traces, the total wave function
$|1 \rangle \oplus |2 \rangle$ is changed to $|2 \rangle \oplus |1
\rangle$ and the total geometric phase is $\pi$. This
is consistent with the anti-symmetry of the fermions.  So we call this
model as the poor man's exchange mode because there are only two
quasi-particles in the two levels system.

In NH system, the trace of the left eigenstates $|n\rangle\rangle$ can be
different from that of the right eigenstates $|n\rangle$. In fig.
\ref{fig1}(b), we enumerate two kinds of the traces, one without loop and
the other with a loop. We numerically calculate the total geometric phase and
find that it is $\pi$ in the former case but can be any
number in the latter case, depending on the details of the traces. We want to emphasize that in the
hermitian case, the geometric phase is still $\pi$ even when there is a
loop in the trace. Here
we only discuss the eigenstates and the NH Hamiltonian in experiment can be build from
them by $\hat H = |1 \rangle E_1 \langle \langle 1 | + |2 \rangle E_2
\langle \langle 2 |$, where $E_i$ are the eigenvalues taking
arbitrary unequal values.

We have found that the total geometric phase becomes path dependent in NH
system.  If one can fix the traces of exchanging two quasi-particles in
this system, the statistics of these quasi-particles will work like
those in anyonic system. Actually in NH model, the eigenstates are not
orthogonal in the sense of traditional inner product, $\langle n |m
\rangle \ne 0$, when $n\ne m$. Such non-orthogonality is also the key ingredient proposed by Haldane in
studying anyonic lattice model\cite{Graner1992}. This property in common makes us surmise that the NH single-particle
model may share many features with anyonic system. In
supporting this assumption, we show that the
effect of ``an action in distance'' can also appear in anyonic system.

Here we will only discuss the quench in anyonic systems in some special
manners. We first use a Lagrangian for anyons\cite{PhysRevB.46.7765}
$L= \sum \frac{m}{2}(\frac{d x^j_i}{dt})^2 +
\frac{\alpha}{\pi}\sum_{r<s}\frac{d\theta(x_r-x_s)}{dt}$, where
$\theta(x_r-x_s)$ is the azimuthal angle between two anyons. If the
many-body interaction are suddenly switched off, the anyonic
system degenerates to the background bosonic or fermionic system and the
interaction represented by
$\theta(x_r -x_s)$ term should disappear. As the Lagrangian includes the
time derivative of $\theta$, the quench will introduce $\delta$-like
functions to the Lagrangian and the wave function should be
discontinuous right at the quench time. 

We can further support the above discussion by investigating an $1D$ anyonic lattice. We
take the hard core anyonic model $H= \sum t_l a^\dagger_{l+1} a_l + {\text
H.C.}$, where $a_l$ and $a^\dagger_l$ are anyonic annihilation and
creation operators\cite{PhysRevA.79.043633}. One can calculate the anyon density by mapping the
Hamiltonian to a fermionic Hamiltonian through a generalized
Jordan-Wigner transformation. The technical details of the calculation
can be found elsewhere\cite{PhysRevA.79.043633}. Here we suppose a
quench start from a uniform chain with $t_l=1$
and suddenly eliminate the hopping at the center $l=N/2$ with $t_{N/2}=0$. After
quench, the system is separated into unrelated two parts and the
density is calculated in these parts individually. Based on the
same fermionic ground state in the fermionic representation,
we calculate the anyonic density and find that the density away from the
source of the quench (the center of the chain) is also changed
immediately after the quench. This indicates that the anyonic wave function must suffer a sudden
change during the quench, which is similar to that in NH
system.

{\it Conclusions and outlooks.}---
We find a TDSE for a general NH system based on several reasonable
ansatzs. Within our theory, the physical quantities are identical and
the formulas are symmetric in the multiple Hilbert spaces spanned by the left eigenstates
and the right eigenstates. As the extra time
derivative metric term explicitly appears in the equation, a quench in
Hamiltonian can cause a discontinuity for the wave functions. This
induces ``an action in distance'', which will violate the no-signaling
condition from special relativity or LRB from non-relativistic quantum
mechanics. It will be interesting to embed the idea of the new TDSE to
Dirac equation to exam whether the no-signaling condition is really
violated in the relativistic NH quantum mechanics. We also find that
the geometric phase of exchanging two quasi-particles is path-dependent
and can be distinguished from $\pi$. So if there is a possibility to fix
the exchanging traces in a special manner, the quasi-particles in NH system should work like
anyons. This makes us propose that the NH single
particle system shares many features with the anyonic system, and it
may be possible to mimic one system with the other in experiments. In
our discussion, we have supposed that the Hamiltonian is not defective.
It will be interesting to extend the study to the defective case in
which the metric is not full rank and its inverse is not well defined.

\bibliographystyle{apsrev}
\bibliography{main}

\begin{thebibliography}{61}
\expandafter\ifx\csname natexlab\endcsname\relax\def\natexlab#1{#1}\fi
\expandafter\ifx\csname bibnamefont\endcsname\relax
  \def\bibnamefont#1{#1}\fi
\expandafter\ifx\csname bibfnamefont\endcsname\relax
  \def\bibfnamefont#1{#1}\fi
\expandafter\ifx\csname citenamefont\endcsname\relax
  \def\citenamefont#1{#1}\fi
\expandafter\ifx\csname url\endcsname\relax
  \def\url#1{\texttt{#1}}\fi
\expandafter\ifx\csname urlprefix\endcsname\relax\def\urlprefix{URL }\fi
\providecommand{\bibinfo}[2]{#2}
\providecommand{\eprint}[2][]{\url{#2}}

\bibitem[{\citenamefont{de~Morisson~Faria and
  Fring}(2006)}]{0305-4470-39-29-018}
\bibinfo{author}{\bibfnamefont{C.~F.} \bibnamefont{de~Morisson~Faria}}
  \bibnamefont{and} \bibinfo{author}{\bibfnamefont{A.}~\bibnamefont{Fring}},
  \bibinfo{journal}{Journal of Physics A: Mathematical and General}
  \textbf{\bibinfo{volume}{39}}, \bibinfo{pages}{9269} (\bibinfo{year}{2006}).

\bibitem[{\citenamefont{Faria and Fring}(2007)}]{Faria2007}
\bibinfo{author}{\bibfnamefont{C.~F.~M.} \bibnamefont{Faria}} \bibnamefont{and}
  \bibinfo{author}{\bibfnamefont{A.}~\bibnamefont{Fring}},
  \bibinfo{journal}{Laser Physics} \textbf{\bibinfo{volume}{17}},
  \bibinfo{pages}{424} (\bibinfo{year}{2007}).

\bibitem[{\citenamefont{Mostafazadeh}(2007)}]{Mostafazadeh2007208}
\bibinfo{author}{\bibfnamefont{A.}~\bibnamefont{Mostafazadeh}},
  \bibinfo{journal}{Physics Letters B} \textbf{\bibinfo{volume}{650}},
  \bibinfo{pages}{208 } (\bibinfo{year}{2007}).

\bibitem[{\citenamefont{Znojil}({\natexlab{a}})}]{Znojil1}
\bibinfo{author}{\bibfnamefont{M.}~\bibnamefont{Znojil}} ,
  \eprint{arXiv:0710.5653}.

\bibitem[{\citenamefont{Mostafazadeh}({\natexlab{a}})}]{Mostafazadeh1}
\bibinfo{author}{\bibfnamefont{A.}~\bibnamefont{Mostafazadeh}}
  , \eprint{arXiv:0711.0137}.

\bibitem[{\citenamefont{Znojil}({\natexlab{b}})}]{Znojil2}
\bibinfo{author}{\bibfnamefont{M.}~\bibnamefont{Znojil}},
  \eprint{arXiv:0711.0514}.

\bibitem[{\citenamefont{Mostafazadeh}({\natexlab{b}})}]{Mostafazadeh2}
\bibinfo{author}{\bibfnamefont{A.}~\bibnamefont{Mostafazadeh}}
  , \eprint{arXiv:0711.1078}.

\bibitem[{\citenamefont{Znojil}(2008)}]{PhysRevD.78.085003}
\bibinfo{author}{\bibfnamefont{M.}~\bibnamefont{Znojil}},
  \bibinfo{journal}{Phys. Rev. D} \textbf{\bibinfo{volume}{78}},
  \bibinfo{pages}{085003} (\bibinfo{year}{2008}).

\bibitem[{\citenamefont{Gong and Wang}(2013)}]{Gong2013}
\bibinfo{author}{\bibfnamefont{J.}~\bibnamefont{Gong}} \bibnamefont{and}
  \bibinfo{author}{\bibfnamefont{Q.-h.} \bibnamefont{Wang}},
  \bibinfo{journal}{Journal of Physics A: Mathematical and Theoretical}
  \textbf{\bibinfo{volume}{46}}, \bibinfo{pages}{485302}
  (\bibinfo{year}{2013}).

\bibitem[{\citenamefont{Maamache}(2015)}]{PhysRevA.92.032106}
\bibinfo{author}{\bibfnamefont{M.}~\bibnamefont{Maamache}},
  \bibinfo{journal}{Phys. Rev. A} \textbf{\bibinfo{volume}{92}},
  \bibinfo{pages}{032106} (\bibinfo{year}{2015}).

\bibitem[{\citenamefont{Fring and
  Moussa}(2016{\natexlab{a}})}]{PhysRevA.93.042114}
\bibinfo{author}{\bibfnamefont{A.}~\bibnamefont{Fring}} \bibnamefont{and}
  \bibinfo{author}{\bibfnamefont{M.~H.~Y.} \bibnamefont{Moussa}},
  \bibinfo{journal}{Phys. Rev. A} \textbf{\bibinfo{volume}{93}},
  \bibinfo{pages}{042114} (\bibinfo{year}{2016}{\natexlab{a}}).

\bibitem[{\citenamefont{Fring and
  Moussa}(2016{\natexlab{b}})}]{PhysRevA.94.042128}
\bibinfo{author}{\bibfnamefont{A.}~\bibnamefont{Fring}} \bibnamefont{and}
  \bibinfo{author}{\bibfnamefont{M.~H.~Y.} \bibnamefont{Moussa}},
  \bibinfo{journal}{Phys. Rev. A} \textbf{\bibinfo{volume}{94}},
  \bibinfo{pages}{042128} (\bibinfo{year}{2016}{\natexlab{b}}).

\bibitem[{\citenamefont{Heisenberg}(1957)}]{RevModPhys.29.269}
\bibinfo{author}{\bibfnamefont{W.}~\bibnamefont{Heisenberg}},
  \bibinfo{journal}{Rev. Mod. Phys.} \textbf{\bibinfo{volume}{29}},
  \bibinfo{pages}{269} (\bibinfo{year}{1957}).

\bibitem[{\citenamefont{Bender and Boettcher}(1998)}]{Bender1998}
\bibinfo{author}{\bibfnamefont{C.~M.} \bibnamefont{Bender}} \bibnamefont{and}
  \bibinfo{author}{\bibfnamefont{S.}~\bibnamefont{Boettcher}},
  \bibinfo{journal}{Physical Review Letters} \textbf{\bibinfo{volume}{80}},
  \bibinfo{pages}{5243} (\bibinfo{year}{1998}).

\bibitem[{\citenamefont{Bender et~al.}(2002)\citenamefont{Bender, Brody, and
  Jones}}]{Bender2002}
\bibinfo{author}{\bibfnamefont{C.~M.} \bibnamefont{Bender}},
  \bibinfo{author}{\bibfnamefont{D.~C.} \bibnamefont{Brody}}, \bibnamefont{and}
  \bibinfo{author}{\bibfnamefont{H.~F.} \bibnamefont{Jones}},
  \bibinfo{journal}{Physical Review Letters} \textbf{\bibinfo{volume}{89}},
  \bibinfo{pages}{270401} (\bibinfo{year}{2002}).

\bibitem[{\citenamefont{Mostafazadeh}(2002{\natexlab{a}})}]{Mostafazadeh2002}
\bibinfo{author}{\bibfnamefont{A.}~\bibnamefont{Mostafazadeh}},
  \bibinfo{journal}{Journal of Mathematical Physics}
  \textbf{\bibinfo{volume}{43}}, \bibinfo{pages}{3944}
  (\bibinfo{year}{2002}{\natexlab{a}}).

\bibitem[{\citenamefont{Cui and Zheng}(2012)}]{Cui2012}
\bibinfo{author}{\bibfnamefont{X.-D.} \bibnamefont{Cui}} \bibnamefont{and}
  \bibinfo{author}{\bibfnamefont{Y.}~\bibnamefont{Zheng}},
  \bibinfo{journal}{Physical Review A} \textbf{\bibinfo{volume}{86}},
  \bibinfo{pages}{064104} (\bibinfo{year}{2012}).

\bibitem[{\citenamefont{Lee et~al.}(2014)\citenamefont{Lee, Hsieh, Flammia, and
  Lee}}]{Lee2014}
\bibinfo{author}{\bibfnamefont{Y.-C.} \bibnamefont{Lee}},
  \bibinfo{author}{\bibfnamefont{M.-h.} \bibnamefont{Hsieh}},
  \bibinfo{author}{\bibfnamefont{S.~T.} \bibnamefont{Flammia}},
  \bibnamefont{and} \bibinfo{author}{\bibfnamefont{R.-k.} \bibnamefont{Lee}},
  \bibinfo{journal}{Physical Review Letters} \textbf{\bibinfo{volume}{112}},
  \bibinfo{pages}{130404} (\bibinfo{year}{2014}).

\bibitem[{\citenamefont{Malzard et~al.}(2015)\citenamefont{Malzard, Poli, and
  Schomerus}}]{Malzard2015}
\bibinfo{author}{\bibfnamefont{S.}~\bibnamefont{Malzard}},
  \bibinfo{author}{\bibfnamefont{C.}~\bibnamefont{Poli}}, \bibnamefont{and}
  \bibinfo{author}{\bibfnamefont{H.}~\bibnamefont{Schomerus}},
  \bibinfo{journal}{Physical Review Letters} \textbf{\bibinfo{volume}{115}},
  \bibinfo{pages}{200402} (\bibinfo{year}{2015}).

\bibitem[{\citenamefont{Medvedyeva et~al.}(2016)\citenamefont{Medvedyeva,
  Essler, and Prosen}}]{Medvedyeva2016}
\bibinfo{author}{\bibfnamefont{M.~V.} \bibnamefont{Medvedyeva}},
  \bibinfo{author}{\bibfnamefont{F.~H.~L.} \bibnamefont{Essler}},
  \bibnamefont{and} \bibinfo{author}{\bibfnamefont{T.}~\bibnamefont{Prosen}},
  \bibinfo{journal}{Physical Review Letters} \textbf{\bibinfo{volume}{117}},
  \bibinfo{pages}{137202} (\bibinfo{year}{2016}).

\bibitem[{\citenamefont{Amir et~al.}(2016)\citenamefont{Amir, Hatano, and
  Nelson}}]{Amir2015}
\bibinfo{author}{\bibfnamefont{A.}~\bibnamefont{Amir}},
  \bibinfo{author}{\bibfnamefont{N.}~\bibnamefont{Hatano}}, \bibnamefont{and}
  \bibinfo{author}{\bibfnamefont{D.~R.} \bibnamefont{Nelson}},
  \bibinfo{journal}{Physical Review E} \textbf{\bibinfo{volume}{93}},
  \bibinfo{pages}{042310} (\bibinfo{year}{2016}).

\bibitem[{\citenamefont{Dattoli et~al.}(1990)\citenamefont{Dattoli, Mignani,
  and Torre}}]{Dattoli1990}
\bibinfo{author}{\bibfnamefont{G.}~\bibnamefont{Dattoli}},
  \bibinfo{author}{\bibfnamefont{R.}~\bibnamefont{Mignani}}, \bibnamefont{and}
  \bibinfo{author}{\bibfnamefont{A.}~\bibnamefont{Torre}},
  \bibinfo{journal}{Journal of Physics A: Mathematical and General}
  \textbf{\bibinfo{volume}{23}}, \bibinfo{pages}{5795} (\bibinfo{year}{1990}).

\bibitem[{\citenamefont{Mehri-Dehnavi and
  Mostafazadeh}(2008)}]{Mehri-Dehnavi2008}
\bibinfo{author}{\bibfnamefont{H.}~\bibnamefont{Mehri-Dehnavi}}
  \bibnamefont{and}
  \bibinfo{author}{\bibfnamefont{A.}~\bibnamefont{Mostafazadeh}},
  \bibinfo{journal}{Journal of Mathematical Physics}
  \textbf{\bibinfo{volume}{49}} (\bibinfo{year}{2008}).

\bibitem[{\citenamefont{M{\"{u}}ller and Rotter}(2009)}]{Muller2009a}
\bibinfo{author}{\bibfnamefont{M.}~\bibnamefont{M{\"{u}}ller}}
  \bibnamefont{and} \bibinfo{author}{\bibfnamefont{I.}~\bibnamefont{Rotter}},
  \bibinfo{journal}{Physical Review A - Atomic, Molecular, and Optical Physics}
  \textbf{\bibinfo{volume}{80}} (\bibinfo{year}{2009}).

\bibitem[{\citenamefont{El-Ganainy et~al.}(2007)\citenamefont{El-Ganainy,
  Makris, Christodoulides, and Musslimani}}]{Gania2010}
\bibinfo{author}{\bibfnamefont{R.}~\bibnamefont{El-Ganainy}},
  \bibinfo{author}{\bibfnamefont{K.~G.} \bibnamefont{Makris}},
  \bibinfo{author}{\bibfnamefont{D.~N.} \bibnamefont{Christodoulides}},
  \bibnamefont{and} \bibinfo{author}{\bibfnamefont{Z.~H.}
  \bibnamefont{Musslimani}}, \bibinfo{journal}{Opt. Lett.}
  \textbf{\bibinfo{volume}{32}}, \bibinfo{pages}{2632} (\bibinfo{year}{2007}).

\bibitem[{\citenamefont{Liang and Huang}(2013)}]{Liang2013}
\bibinfo{author}{\bibfnamefont{S.-D.} \bibnamefont{Liang}} \bibnamefont{and}
  \bibinfo{author}{\bibfnamefont{G.-Y.} \bibnamefont{Huang}},
  \bibinfo{journal}{Physical Review A} \textbf{\bibinfo{volume}{87}},
  \bibinfo{pages}{012118} (\bibinfo{year}{2013}).

\bibitem[{\citenamefont{Cui and Zheng}(2014)}]{Cui2014a}
\bibinfo{author}{\bibfnamefont{X.-D.} \bibnamefont{Cui}} \bibnamefont{and}
  \bibinfo{author}{\bibfnamefont{Y.}~\bibnamefont{Zheng}},
  \bibinfo{journal}{Scientific Reports} \textbf{\bibinfo{volume}{4}},
  \bibinfo{pages}{5813} (\bibinfo{year}{2014}).

\bibitem[{\citenamefont{Lee and Chan}(2014)}]{Lee2014b}
\bibinfo{author}{\bibfnamefont{T.~E.} \bibnamefont{Lee}} \bibnamefont{and}
  \bibinfo{author}{\bibfnamefont{C.-k.} \bibnamefont{Chan}},
  \bibinfo{journal}{Physical Review X} \textbf{\bibinfo{volume}{4}},
  \bibinfo{pages}{041001} (\bibinfo{year}{2014}).

\bibitem[{\citenamefont{Bender et~al.}(2014)\citenamefont{Bender, Hook,
  Mavromatos, and Sarkar}}]{Bender2014a}
\bibinfo{author}{\bibfnamefont{C.~M.} \bibnamefont{Bender}},
  \bibinfo{author}{\bibfnamefont{D.~W.} \bibnamefont{Hook}},
  \bibinfo{author}{\bibfnamefont{N.~E.} \bibnamefont{Mavromatos}},
  \bibnamefont{and} \bibinfo{author}{\bibfnamefont{S.}~\bibnamefont{Sarkar}},
  \bibinfo{journal}{Physical Review Letters} \textbf{\bibinfo{volume}{113}},
  \bibinfo{pages}{231605} (\bibinfo{year}{2014}).

\bibitem[{\citenamefont{Shah et~al.}(2015)\citenamefont{Shah, Chattopadhyay,
  Vaidya, and Chakraborty}}]{Shah2015}
\bibinfo{author}{\bibfnamefont{T.}~\bibnamefont{Shah}},
  \bibinfo{author}{\bibfnamefont{R.}~\bibnamefont{Chattopadhyay}},
  \bibinfo{author}{\bibfnamefont{K.}~\bibnamefont{Vaidya}}, \bibnamefont{and}
  \bibinfo{author}{\bibfnamefont{S.}~\bibnamefont{Chakraborty}},
  \bibinfo{journal}{Physical Review E} \textbf{\bibinfo{volume}{92}},
  \bibinfo{pages}{062927} (\bibinfo{year}{2015}).

\bibitem[{\citenamefont{Nixon and Yang}(2016)}]{Nixon2015}
\bibinfo{author}{\bibfnamefont{S.}~\bibnamefont{Nixon}} \bibnamefont{and}
  \bibinfo{author}{\bibfnamefont{J.}~\bibnamefont{Yang}},
  \bibinfo{journal}{Physical Review A} \textbf{\bibinfo{volume}{93}},
  \bibinfo{pages}{031802} (\bibinfo{year}{2016}).

\bibitem[{\citenamefont{Ye et~al.}()\citenamefont{Ye, Tianxiang, Xiaohui, and
  Peiqing}}]{Xiong1}
\bibinfo{author}{\bibfnamefont{X.}~\bibnamefont{Ye}},
  \bibinfo{author}{\bibfnamefont{W.}~\bibnamefont{Tianxiang}},
  \bibinfo{author}{\bibfnamefont{W.}~\bibnamefont{Xiaohui}}, \bibnamefont{and}
  \bibinfo{author}{\bibfnamefont{T.}~\bibnamefont{Peiqing}},
  \eprint{arXiv:1610.06275}.

\bibitem[{\citenamefont{Ruschhaupt et~al.}(2005)\citenamefont{Ruschhaupt,
  Delgado, and Muga}}]{0305-4470-38-9-L03}
\bibinfo{author}{\bibfnamefont{A.}~\bibnamefont{Ruschhaupt}},
  \bibinfo{author}{\bibfnamefont{F.}~\bibnamefont{Delgado}}, \bibnamefont{and}
  \bibinfo{author}{\bibfnamefont{J.~G.} \bibnamefont{Muga}},
  \bibinfo{journal}{Journal of Physics A: Mathematical and General}
  \textbf{\bibinfo{volume}{38}}, \bibinfo{pages}{L171} (\bibinfo{year}{2005}).

\bibitem[{\citenamefont{Makris et~al.}(2008)\citenamefont{Makris, El-Ganainy,
  Christodoulides, and Musslimani}}]{PhysRevLett.100.103904}
\bibinfo{author}{\bibfnamefont{K.~G.} \bibnamefont{Makris}},
  \bibinfo{author}{\bibfnamefont{R.}~\bibnamefont{El-Ganainy}},
  \bibinfo{author}{\bibfnamefont{D.~N.} \bibnamefont{Christodoulides}},
  \bibnamefont{and} \bibinfo{author}{\bibfnamefont{Z.~H.}
  \bibnamefont{Musslimani}}, \bibinfo{journal}{Phys. Rev. Lett.}
  \textbf{\bibinfo{volume}{100}}, \bibinfo{pages}{103904}
  (\bibinfo{year}{2008}).

\bibitem[{\citenamefont{Guo et~al.}(2009)\citenamefont{Guo, Salamo, Duchesne,
  Morandotti, Volatier-Ravat, Aimez, Siviloglou, and
  Christodoulides}}]{PhysRevLett.103.093902}
\bibinfo{author}{\bibfnamefont{A.}~\bibnamefont{Guo}},
  \bibinfo{author}{\bibfnamefont{G.~J.} \bibnamefont{Salamo}},
  \bibinfo{author}{\bibfnamefont{D.}~\bibnamefont{Duchesne}},
  \bibinfo{author}{\bibfnamefont{R.}~\bibnamefont{Morandotti}},
  \bibinfo{author}{\bibfnamefont{M.}~\bibnamefont{Volatier-Ravat}},
  \bibinfo{author}{\bibfnamefont{V.}~\bibnamefont{Aimez}},
  \bibinfo{author}{\bibfnamefont{G.~A.} \bibnamefont{Siviloglou}},
  \bibnamefont{and} \bibinfo{author}{\bibfnamefont{D.~N.}
  \bibnamefont{Christodoulides}}, \bibinfo{journal}{Phys. Rev. Lett.}
  \textbf{\bibinfo{volume}{103}}, \bibinfo{pages}{093902}
  (\bibinfo{year}{2009}).

\bibitem[{\citenamefont{Longhi}(2010)}]{PhysRevA.82.031801}
\bibinfo{author}{\bibfnamefont{S.}~\bibnamefont{Longhi}},
  \bibinfo{journal}{Phys. Rev. A} \textbf{\bibinfo{volume}{82}},
  \bibinfo{pages}{031801} (\bibinfo{year}{2010}).

\bibitem[{\citenamefont{Longhi}(2009)}]{PhysRevLett.103.123601}
\bibinfo{author}{\bibfnamefont{S.}~\bibnamefont{Longhi}},
  \bibinfo{journal}{Phys. Rev. Lett.} \textbf{\bibinfo{volume}{103}},
  \bibinfo{pages}{123601} (\bibinfo{year}{2009}).

\bibitem[{\citenamefont{R{\"{u}}ter et~al.}(2010)\citenamefont{R{\"{u}}ter,
  Makris, El-Ganainy, Christodoulides, Segev, and Kip}}]{Ruter2010}
\bibinfo{author}{\bibfnamefont{C.~E.} \bibnamefont{R{\"{u}}ter}},
  \bibinfo{author}{\bibfnamefont{K.~G.} \bibnamefont{Makris}},
  \bibinfo{author}{\bibfnamefont{R.}~\bibnamefont{El-Ganainy}},
  \bibinfo{author}{\bibfnamefont{D.~N.} \bibnamefont{Christodoulides}},
  \bibinfo{author}{\bibfnamefont{M.}~\bibnamefont{Segev}}, \bibnamefont{and}
  \bibinfo{author}{\bibfnamefont{D.}~\bibnamefont{Kip}},
  \bibinfo{journal}{Nature Physics} \textbf{\bibinfo{volume}{6}},
  \bibinfo{pages}{192} (\bibinfo{year}{2010}).

\bibitem[{\citenamefont{Ramezani et~al.}(2010)\citenamefont{Ramezani, Kottos,
  El-Ganainy, and Christodoulides}}]{PhysRevA.82.043803}
\bibinfo{author}{\bibfnamefont{H.}~\bibnamefont{Ramezani}},
  \bibinfo{author}{\bibfnamefont{T.}~\bibnamefont{Kottos}},
  \bibinfo{author}{\bibfnamefont{R.}~\bibnamefont{El-Ganainy}},
  \bibnamefont{and} \bibinfo{author}{\bibfnamefont{D.~N.}
  \bibnamefont{Christodoulides}}, \bibinfo{journal}{Phys. Rev. A}
  \textbf{\bibinfo{volume}{82}}, \bibinfo{pages}{043803}
  (\bibinfo{year}{2010}).

\bibitem[{\citenamefont{Lin et~al.}(2011)\citenamefont{Lin, Ramezani,
  Eichelkraut, Kottos, Cao, and Christodoulides}}]{PhysRevLett.106.213901}
\bibinfo{author}{\bibfnamefont{Z.}~\bibnamefont{Lin}},
  \bibinfo{author}{\bibfnamefont{H.}~\bibnamefont{Ramezani}},
  \bibinfo{author}{\bibfnamefont{T.}~\bibnamefont{Eichelkraut}},
  \bibinfo{author}{\bibfnamefont{T.}~\bibnamefont{Kottos}},
  \bibinfo{author}{\bibfnamefont{H.}~\bibnamefont{Cao}}, \bibnamefont{and}
  \bibinfo{author}{\bibfnamefont{D.~N.} \bibnamefont{Christodoulides}},
  \bibinfo{journal}{Phys. Rev. Lett.} \textbf{\bibinfo{volume}{106}},
  \bibinfo{pages}{213901} (\bibinfo{year}{2011}).

\bibitem[{\citenamefont{Lin et~al.}(2012)\citenamefont{Lin, Schindler, Ellis,
  and Kottos}}]{PhysRevA.85.050101}
\bibinfo{author}{\bibfnamefont{Z.}~\bibnamefont{Lin}},
  \bibinfo{author}{\bibfnamefont{J.}~\bibnamefont{Schindler}},
  \bibinfo{author}{\bibfnamefont{F.~M.} \bibnamefont{Ellis}}, \bibnamefont{and}
  \bibinfo{author}{\bibfnamefont{T.}~\bibnamefont{Kottos}},
  \bibinfo{journal}{Phys. Rev. A} \textbf{\bibinfo{volume}{85}},
  \bibinfo{pages}{050101} (\bibinfo{year}{2012}).

\bibitem[{\citenamefont{Bender et~al.}(2013)\citenamefont{Bender, Factor,
  Bodyfelt, Ramezani, Christodoulides, Ellis, and
  Kottos}}]{PhysRevLett.110.234101}
\bibinfo{author}{\bibfnamefont{N.}~\bibnamefont{Bender}},
  \bibinfo{author}{\bibfnamefont{S.}~\bibnamefont{Factor}},
  \bibinfo{author}{\bibfnamefont{J.~D.} \bibnamefont{Bodyfelt}},
  \bibinfo{author}{\bibfnamefont{H.}~\bibnamefont{Ramezani}},
  \bibinfo{author}{\bibfnamefont{D.~N.} \bibnamefont{Christodoulides}},
  \bibinfo{author}{\bibfnamefont{F.~M.} \bibnamefont{Ellis}}, \bibnamefont{and}
  \bibinfo{author}{\bibfnamefont{T.}~\bibnamefont{Kottos}},
  \bibinfo{journal}{Phys. Rev. Lett.} \textbf{\bibinfo{volume}{110}},
  \bibinfo{pages}{234101} (\bibinfo{year}{2013}).

\bibitem[{\citenamefont{Fleury et~al.}(2014)\citenamefont{Fleury, Sounas, and
  Al\`u}}]{PhysRevLett.113.023903}
\bibinfo{author}{\bibfnamefont{R.}~\bibnamefont{Fleury}},
  \bibinfo{author}{\bibfnamefont{D.~L.} \bibnamefont{Sounas}},
  \bibnamefont{and} \bibinfo{author}{\bibfnamefont{A.}~\bibnamefont{Al\`u}},
  \bibinfo{journal}{Phys. Rev. Lett.} \textbf{\bibinfo{volume}{113}},
  \bibinfo{pages}{023903} (\bibinfo{year}{2014}).

\bibitem[{\citenamefont{Regensburger et~al.}(2013)\citenamefont{Regensburger,
  Miri, Bersch, N{\"{a}}ger, Onishchukov, Christodoulides, and
  Peschel}}]{Regensburger2013}
\bibinfo{author}{\bibfnamefont{A.}~\bibnamefont{Regensburger}},
  \bibinfo{author}{\bibfnamefont{M.-A.} \bibnamefont{Miri}},
  \bibinfo{author}{\bibfnamefont{C.}~\bibnamefont{Bersch}},
  \bibinfo{author}{\bibfnamefont{J.}~\bibnamefont{N{\"{a}}ger}},
  \bibinfo{author}{\bibfnamefont{G.}~\bibnamefont{Onishchukov}},
  \bibinfo{author}{\bibfnamefont{D.~N.} \bibnamefont{Christodoulides}},
  \bibnamefont{and} \bibinfo{author}{\bibfnamefont{U.}~\bibnamefont{Peschel}},
  \bibinfo{journal}{Physical Review Letters} \textbf{\bibinfo{volume}{110}},
  \bibinfo{pages}{223902} (\bibinfo{year}{2013}).

\bibitem[{\citenamefont{Zhang et~al.}(2016)\citenamefont{Zhang, Zhang, Sheng,
  Yang, Miri, Christodoulides, He, Zhang, and Xiao}}]{Zhang2016g}
\bibinfo{author}{\bibfnamefont{Z.}~\bibnamefont{Zhang}},
  \bibinfo{author}{\bibfnamefont{Y.}~\bibnamefont{Zhang}},
  \bibinfo{author}{\bibfnamefont{J.}~\bibnamefont{Sheng}},
  \bibinfo{author}{\bibfnamefont{L.}~\bibnamefont{Yang}},
  \bibinfo{author}{\bibfnamefont{M.-a.} \bibnamefont{Miri}},
  \bibinfo{author}{\bibfnamefont{D.~N.} \bibnamefont{Christodoulides}},
  \bibinfo{author}{\bibfnamefont{B.}~\bibnamefont{He}},
  \bibinfo{author}{\bibfnamefont{Y.}~\bibnamefont{Zhang}}, \bibnamefont{and}
  \bibinfo{author}{\bibfnamefont{M.}~\bibnamefont{Xiao}},
  \bibinfo{journal}{Physical Review Letters} \textbf{\bibinfo{volume}{117}},
  \bibinfo{pages}{123601} (\bibinfo{year}{2016}).

\bibitem[{\citenamefont{Hodaei et~al.}(2014)\citenamefont{Hodaei, Miri,
  Heinrich, Christodoulides, and Khajavikhan}}]{Hodaei2014}
\bibinfo{author}{\bibfnamefont{H.}~\bibnamefont{Hodaei}},
  \bibinfo{author}{\bibfnamefont{M.-a.} \bibnamefont{Miri}},
  \bibinfo{author}{\bibfnamefont{M.}~\bibnamefont{Heinrich}},
  \bibinfo{author}{\bibfnamefont{D.~N.} \bibnamefont{Christodoulides}},
  \bibnamefont{and}
  \bibinfo{author}{\bibfnamefont{M.}~\bibnamefont{Khajavikhan}},
  \bibinfo{journal}{Science} \textbf{\bibinfo{volume}{975}}
  (\bibinfo{year}{2014}).

\bibitem[{\citenamefont{Zhu et~al.}(2014)\citenamefont{Zhu, Ramezani, Shi, Zhu,
  and Zhang}}]{PhysRevX.4.031042}
\bibinfo{author}{\bibfnamefont{X.}~\bibnamefont{Zhu}},
  \bibinfo{author}{\bibfnamefont{H.}~\bibnamefont{Ramezani}},
  \bibinfo{author}{\bibfnamefont{C.}~\bibnamefont{Shi}},
  \bibinfo{author}{\bibfnamefont{J.}~\bibnamefont{Zhu}}, \bibnamefont{and}
  \bibinfo{author}{\bibfnamefont{X.}~\bibnamefont{Zhang}},
  \bibinfo{journal}{Phys. Rev. X} \textbf{\bibinfo{volume}{4}},
  \bibinfo{pages}{031042} (\bibinfo{year}{2014}).

\bibitem[{\citenamefont{Fleury et~al.}(2015)\citenamefont{Fleury, Sounas, and
  Al{\`{u}}}}]{Fleury2015}
\bibinfo{author}{\bibfnamefont{R.}~\bibnamefont{Fleury}},
  \bibinfo{author}{\bibfnamefont{D.}~\bibnamefont{Sounas}}, \bibnamefont{and}
  \bibinfo{author}{\bibfnamefont{A.}~\bibnamefont{Al{\`{u}}}},
  \bibinfo{journal}{Nature Communications} \textbf{\bibinfo{volume}{6}},
  \bibinfo{pages}{5905} (\bibinfo{year}{2015}).

\bibitem[{\citenamefont{Chen and Jung}(2016)}]{Chen2016}
\bibinfo{author}{\bibfnamefont{P.-Y.} \bibnamefont{Chen}} \bibnamefont{and}
  \bibinfo{author}{\bibfnamefont{J.}~\bibnamefont{Jung}},
  \bibinfo{journal}{Physical Review Applied} \textbf{\bibinfo{volume}{5}},
  \bibinfo{pages}{064018} (\bibinfo{year}{2016}).

\bibitem[{\citenamefont{Scholtz et~al.}(1992)\citenamefont{Scholtz, Geyer, and
  Hahne}}]{Scholtz1992}
\bibinfo{author}{\bibfnamefont{F.}~\bibnamefont{Scholtz}},
  \bibinfo{author}{\bibfnamefont{H.~B.} \bibnamefont{Geyer}}, \bibnamefont{and}
  \bibinfo{author}{\bibfnamefont{F.~J.~W.} \bibnamefont{Hahne}},
  \bibinfo{journal}{Ann. Phys. (NY)} \textbf{\bibinfo{volume}{213}},
  \bibinfo{pages}{74} (\bibinfo{year}{1992}).

\bibitem[{\citenamefont{Mostafazadeh}(2003)}]{Mocal3}
\bibinfo{author}{\bibfnamefont{A.}~\bibnamefont{Mostafazadeh}},
  \bibinfo{journal}{Journal of Physics A} \textbf{\bibinfo{volume}{36}},
  \bibinfo{pages}{7081} (\bibinfo{year}{2003}).

\bibitem[{\citenamefont{Czech}(2004)}]{Czech1}
\bibinfo{author}{\bibnamefont{Czech}}, \bibinfo{journal}{J. Phys.}
  \textbf{\bibinfo{volume}{54}}, \bibinfo{pages}{1125} (\bibinfo{year}{2004}).

\bibitem[{\citenamefont{Mostafazadeh and Batal}(2004)}]{Mocal4}
\bibinfo{author}{\bibfnamefont{A.}~\bibnamefont{Mostafazadeh}}
  \bibnamefont{and} \bibinfo{author}{\bibfnamefont{A.}~\bibnamefont{Batal}},
  \bibinfo{journal}{Journal of Physics A} \textbf{\bibinfo{volume}{37}},
  \bibinfo{pages}{11645} (\bibinfo{year}{2004}).

\bibitem[{\citenamefont{Mostafazadeh}(2005)}]{Mocal5}
\bibinfo{author}{\bibfnamefont{A.}~\bibnamefont{Mostafazadeh}},
  \bibinfo{journal}{Journal of Physics A} \textbf{\bibinfo{volume}{38}},
  \bibinfo{pages}{3213} (\bibinfo{year}{2005}).

\bibitem[{\citenamefont{Laughlin}(1983)}]{Laughlin1983}
\bibinfo{author}{\bibfnamefont{R.~B.} \bibnamefont{Laughlin}},
  \bibinfo{journal}{Physical Review Letters} \textbf{\bibinfo{volume}{50}},
  \bibinfo{pages}{1395} (\bibinfo{year}{1983}).

\bibitem[{\citenamefont{Arovas et~al.}(1984)\citenamefont{Arovas, Schrieffer,
  and Wilczek}}]{Arovas1984}
\bibinfo{author}{\bibfnamefont{D.}~\bibnamefont{Arovas}},
  \bibinfo{author}{\bibfnamefont{J.}~\bibnamefont{Schrieffer}},
  \bibnamefont{and} \bibinfo{author}{\bibfnamefont{F.}~\bibnamefont{Wilczek}},
  \bibinfo{journal}{Physical Review Letters} \textbf{\bibinfo{volume}{53}},
  \bibinfo{pages}{722} (\bibinfo{year}{1984}).

\bibitem[{\citenamefont{Haldane}(1991)}]{Graner1992}
\bibinfo{author}{\bibfnamefont{F.~D.~M.} \bibnamefont{Haldane}},
  \bibinfo{journal}{Physical Review Letters} \textbf{\bibinfo{volume}{67}},
  \bibinfo{pages}{937} (\bibinfo{year}{1991}).

\bibitem[{\citenamefont{Wieser}(2013)}]{Wieser2013}
\bibinfo{author}{\bibfnamefont{R.}~\bibnamefont{Wieser}},
  \bibinfo{journal}{Physical Review Letters} \textbf{\bibinfo{volume}{110}},
  \bibinfo{pages}{147201} (\bibinfo{year}{2013}).

\bibitem[{\citenamefont{Mostafazadeh}(2002{\natexlab{b}})}]{Mocal6}
\bibinfo{author}{\bibfnamefont{A.}~\bibnamefont{Mostafazadeh}},
  \bibinfo{journal}{Journal of Physics A} \textbf{\bibinfo{volume}{43}},
  \bibinfo{pages}{2814} (\bibinfo{year}{2002}{\natexlab{b}}).

\bibitem[{\citenamefont{Ezawa et~al.}(1992)\citenamefont{Ezawa, Hotta, and
  Iwazaki}}]{PhysRevB.46.7765}
\bibinfo{author}{\bibfnamefont{Z.~F.} \bibnamefont{Ezawa}},
  \bibinfo{author}{\bibfnamefont{M.}~\bibnamefont{Hotta}}, \bibnamefont{and}
  \bibinfo{author}{\bibfnamefont{A.}~\bibnamefont{Iwazaki}},
  \bibinfo{journal}{Phys. Rev. B} \textbf{\bibinfo{volume}{46}},
  \bibinfo{pages}{7765} (\bibinfo{year}{1992}).

\bibitem[{\citenamefont{Hao et~al.}(2009)\citenamefont{Hao, Zhang, and
  Chen}}]{PhysRevA.79.043633}
\bibinfo{author}{\bibfnamefont{Y.}~\bibnamefont{Hao}},
  \bibinfo{author}{\bibfnamefont{Y.}~\bibnamefont{Zhang}}, \bibnamefont{and}
  \bibinfo{author}{\bibfnamefont{S.}~\bibnamefont{Chen}},
  \bibinfo{journal}{Phys. Rev. A} \textbf{\bibinfo{volume}{79}},
  \bibinfo{pages}{043633} (\bibinfo{year}{2009}).

\end{thebibliography}

\end{document}